\documentclass[12pt]{article}
\usepackage{amsfonts,graphicx,amsmath,amsthm,amssymb,epsfig}
\usepackage{float}
\allowdisplaybreaks
\begin{document}

\title{\bf Influence of $f(\mathcal{R},\mathcal{T},\mathcal{Q})$ Gravity on Cylindrical Collapse}
\author{M. Sharif$^1$ \thanks{msharif.math@pu.edu.pk}~ and
Tayyab Naseer$^{1,2}$ \thanks{tayyabnaseer48@yahoo.com; tayyab.naseer@math.uol.edu.pk}\\
$^1$ Department of Mathematics and Statistics, The University of Lahore,\\
1-KM Defence Road Lahore, Pakistan.\\
$^2$ Department of Mathematics, University of the Punjab,\\
Quaid-i-Azam Campus, Lahore-54590, Pakistan.}

\date{}
\maketitle

\begin{abstract}
This article examines the dynamics of gravitational collapse in
$f(\mathcal{R},\mathcal{T},\mathcal{Q})$ gravity, where
$\mathcal{Q}=\mathcal{R}_{\mathrm{ab}}\mathcal{T}^{\mathrm{ab}}$. We
consider self-gravitating anisotropic cylindrical geometry whose
interior is filled with dissipative matter configuration and match
it with exterior cylindrically symmetric spacetime at the
hypersurface through junction conditions. We employ the Misner-Sharp
and M\"{u}ler-Israel Stewart formalisms to derive the dynamical as
well as transport equations corresponding to the model
$\mathcal{R}+\Phi\sqrt{\mathcal{T}}+\Psi\mathcal{Q}$, where $\Phi$
and $\Psi$ are arbitrary coupling constants. We then establish some
relations between these equations through which the impact of
effective matter variables, heat dissipation and the bulk viscosity
on the collapse rate is studied. Further, we express the Weyl scalar
in terms of the effective matter sector. We also obtain the
conformal flatness by applying some restrictions on the considered
model and taking dust configuration into the account. Finally, we
investigate various cases to check whether the modified corrections
increase or decrease the collapse rate.
\end{abstract}

\section{Introduction}

Cosmological observations reveal that our universe is originated by
the expansion of superheated matter and energy. Cosmologists
explored that a considerable portion of this unfathomable universe
is made up of stars, planets and galaxies. The most appealing and
promising phenomenon in the structural formation of these celestial
objects is the gravitational collapse. The pioneer work of
Chandrasekhar \cite{28} on this phenomenon helps scientists to
understand its importance in the field of relativistic astrophysics.
He found that a star remains stable until its external pressure and
internal force of attraction (due to its mass) are counterbalanced
by each other. The dynamics of dust collapse has been discussed by
Oppenheimer and Snyder \cite{29}, from which they found that such
collapse eventually results in a black hole. Misner and Sharp
\cite{29a} studied spherical geometry coupled with anisotropic fluid
and checked how the collapse rate is affected by pressure
anisotropy. The Misner-Sharp technique has been employed by Herrera
and Santos \cite{30} to investigate the collapsing rate of a sphere
and found that the energy dissipates in the form of heat/radiations.
Herrera et al. \cite{31} analyzed the impact of anisotropy on the
collapse of cylindrically symmetric matter source. Sharif and his
collaborators \cite{32} studied the dynamics of uncharged and
charged spherical/cylindrical systems and deduced that the collapse
rate is reduced in the presence of electric charge.

As the process of gravitational collapse is highly dissipative, the
effects of heat dissipation in this phenomenon cannot be ignored
\cite{32a,33}. Chan \cite{34} explored the collapsing phenomenon for
a radiating compact object and revealed that the shear viscosity
increases anisotropy of the fluid distribution. Di Prisco \emph{et
al.} \cite{35c} studied anisotropic matter configuration and
disclosed that the explosion in the internal region of spherical
geometry causes the formation of singularity. Nath et al. \cite{35a}
examined the collapsing rate by employing matching criteria between
quasi-spherical Szekeres and charged Vaidya spacetimes as interior
and exterior geometries, respectively. They concluded that the
formation of naked singularity is supported by electric charge.
Herrera et al. \cite{35} discussed self-gravitating viscous
dissipative fluid and found that the dissipative parameters decrease
the force of gravity that eventually decreases the collapsing rate.

Cylindrical gravitational waves exist and support cylindrically
symmetric self-gravitating structures whose study yields significant
consequences. Such geometrical objects prompted many researchers to
explore their different fundamental characteristics. The study of
these structures was pioneered by Bronnikov and Kovalchuk
\cite{35d}. Wang \cite{35e} studied four-dimensional cylinder and
determined exact solutions to the field equations corresponding to a
massless scalar field. He found that collapse of such object may
result in the black hole. Guha and Banerji \cite{36} studied the
dynamics of cylindrical anisotropic geometry, experiencing heat
dissipation and undergoing the gravitational collapse, and derived
the solutions for the matter source. In stellar evolution, the Weyl
tensor plays a significant role that helps to measure the curvature
of geometrical structure. The gravitational collapse of a sphere has
been discussed by Penrose \cite{36a} by formulating a relation
between state variables and the Weyl tensor. Sharif and Fatima
\cite{36b} represented the Weyl tensor in terms of matter variables,
anisotropy and coefficient of shear viscosity for cylindrical
configuration, and discussed the collapsing rate. They found
conformal flatness condition corresponding to the homogeneous energy
density.

The current accelerated expansion of the universe is considered as
the most fascinating phenomenon in the field of cosmology and
astrophysics for the last couple of years \cite{1a,1b}. This
expansion was claimed to be triggered by an obscure form of force
having immense repulsive effects, known as dark energy. The study of
such cosmic nature in the theory of general relativity
($\mathbb{GR}$) faces some deficiencies like cosmic coincidence and
fine-tuning probelm. In this context, scientists developed several
modifications to $\mathbb{GR}$ to address such issues appropriately.
To study cosmological outcomes at large scales, the simplest
possible generalization of $\mathbb{GR}$ was attained by replacing
the Ricci scalar with its generic functional, named $f(\mathcal{R})$
theory \cite{1c}. Different modified $f(\mathcal{R})$ models have
been investigated through multiple approaches and the obtained
results are found to be physically feasible \cite{2}-\cite{2d}.

The idea of matter-geometry coupling was initially presented by
Bertolami et al. \cite{10} to study the appealing nature as well as
composition of the universe. They analyzed the impact of such
interaction in $f(\mathcal{R})$ framework by engaging the
geometrical term in the fluid Lagrangian. This interaction was
recently generalized by Harko et al. \cite{20} at action level, who
introduced $f(\mathcal{R},\mathcal{T})$ gravity, $\mathcal{T}$ being
trace of the energy-momentum tensor $(\mathrm{EMT})$. The
gravitational theories involving such a matter term results in its
non-vanishing divergence. This theory produces several astonishing
results corresponding to self-gravitating structures
\cite{21a}-\cite{21e}. However, the $f(\mathcal{R},\mathcal{T})$
gravity fails to entail the coupling effects on compact bodies at
some point, thus one needs to overcome this issue. Haghani et al.
\cite{22}, in this regard, generalized the functional by inserting
an additional term $\mathcal{Q}$, that represents contraction of
$\mathrm{EMT}$ with the Ricci tensor $\mathcal{R}_{\mathrm{ab}}$.
They considered three different models in
$f(\mathcal{R},\mathcal{T},\mathcal{Q})$ gravity and studied their
respective cosmological applications for high density regime as well
as pressureless matter fluid case.

Sharif and Zubair \cite{22a} studied the thermodynamical laws for
the black hole by adopting the models
$\mathcal{R}+\lambda\mathcal{Q}$ and
$\mathcal{R}(1+\lambda\mathcal{Q})$ along with matter Lagrangian in
terms of energy density as well as pressure. The energy bounds are
also addressed in this scenario, from which they concluded that only
positive values of the model parameter $\lambda$ satisfy weak energy
conditions \cite{22b}. The flat FLRW spacetime was considered to
check the behavior of this extended theory for different
cosmological models \cite{23}. They reconstructed the modified
gravitational action and also described de Sitter universe solutions
corresponding to the perfect fluid distribution. Baffou et al.
\cite{25} performed stability analysis in this modified gravity for
two particular cases and concluded that both models present
stability through some perturbation functions. Yousaf et al.
\cite{26} performed orthogonal decomposition of the Riemann tensor
on the effective $\mathrm{EMT}$ corresponding to static/non-static
spherical structures and computed some scalars to discuss the
structural evolution of these bodies. The evolutionary patterns for
cylindrical spacetime have also been discussed in modified scenario
\cite{26a}. We also studied charged/uncharged sphere and obtained
several physically acceptable anisotropic solutions through
different schemes \cite{27,27aa}.

The extensive discussion on the collapsing phenomenon has been done
in various modified backgrounds. The numerical simulations were
employed to study the collapse of spherical body in $f(\mathcal{R})$
framework from which an unusual increment in the density of fluid
has been found \cite{27ab}. In this context, Shamir and Fayyaz
\cite{27ac} analyzed the dynamics of a self-gravitational
cylindrical geometry whose interior is filled with anisotropic
dissipative fluid. They found the collapse as the crucial element to
examine the rapid acceleration. The dynamical equations have been
used to investigate the behavior of anisotropic gravitating source
in $f(\mathcal{R},\mathcal{T})$ scenario \cite{27ad}. Bhatti
\emph{et al.} \cite{27aad} discussed the collapsing rate of
dissipative anisotropic matter distribution with/without involving
the effects of radiation density and coefficient of shear viscosity
in $f(\mathcal{R},\mathcal{T},\mathcal{Q})$ framework. Sharif et al.
\cite{27af} studied the celestial objects coupled with
perfect/anisotropic configurations with/without the heat dissipation
in different modified theories, and examined the influence of
correction terms on the collapsing rate.

This article addresses the dynamics of cylindrical geometry
involving the impact of principal stresses and the dissipation flux
in
$f(\mathcal{R},\mathcal{T},\mathcal{R}_{\mathrm{ab}}\mathcal{T}^{\mathrm{ab}})$
theory. The paper is structured in the following format. We define
some elementary terms related to the collapse and the extended
theory, and formulate the corresponding equations of motion and
non-vanishing dynamical identities for
$\mathcal{R}+\Phi\sqrt{\mathcal{T}}+\Psi\mathcal{Q}$ in section
\textbf{2}. Moreover, the C-energy and the junction conditions are
calculated through Darmois criteria. Section \textbf{3} formulates
the dynamical equations and then couple them with the acceleration
of the fluid. We further construct several dynamical forces in
modified gravity in section \textbf{4} to study their impact on the
collapsing rate. Section \textbf{5} explores some interesting
relations between the effective state parameters and the Weyl
scalar. The last section summarizes all of our findings.

\section{$f(\mathcal{R},\mathcal{T},\mathcal{R}_{\mathrm{a}\mathrm{b}}\mathcal{T}^{\mathrm{a}\mathrm{b}})$ Theory}

The modified Einstein-Hilbert action for the
$f(\mathcal{R},\mathcal{T},\mathcal{Q})$ gravity (with
$\kappa=8\pi$) has the following form \cite{23}
\begin{equation}\label{g1}
\mathbb{S}=\int\left\{\frac{f(\mathcal{R},\mathcal{T},\mathcal{Q})}{16\pi}
+\mathbb{L}_{\mathcal{M}}\right\}\sqrt{-g}d^{4}x,
\end{equation}
where $\mathbb{L}_{\mathcal{M}}$ is the Lagrangian corresponding to
the matter density. Implementing the principle of least-action on
Eq.\eqref{g1} provides
\begin{equation}\label{g2}
\mathcal{G}_{\mathrm{a}\mathrm{b}}=\mathcal{T}_{\mathrm{a}\mathrm{b}}^{(\mathrm{EFF})}=\frac{1}
{f_{\mathcal{R}}-\mathbb{L}_{\mathcal{M}}f_{\mathcal{Q}}}\left(8\pi\mathcal{T}_{\mathrm{a}\mathrm{b}}+\mathcal{T}_{\mathrm{a}\mathrm{b}}^{(D)}\right),
\end{equation}
where, $\mathcal{T}_{\mathrm{a}\mathrm{b}}^{(\mathrm{EFF})}$ and
$\mathcal{T}_{\mathrm{a}\mathrm{b}}$ are termed as the effective and
usual anisotropic matter $\mathrm{EMT}$. Also,
$\mathcal{G}_{\mathrm{a}\mathrm{b}}$ is the Einstein tensor. The
modified corrections are represented by
$\mathcal{T}_{\mathrm{a}\mathrm{b}}^{(D)}$ which has the form
\begin{eqnarray}\nonumber
\mathcal{T}_{\mathrm{a}\mathrm{b}}^{(D)}&=&\left(f_{\mathcal{T}}+\frac{1}{2}\mathcal{R}f_{\mathcal{Q}}\right)\mathcal{T}_{\mathrm{a}\mathrm{b}}
+\left\{\frac{\mathcal{R}}{2}\left(\frac{f}{\mathcal{R}}-f_{\mathcal{R}}\right)
-\frac{1}{2}\nabla_{\mathrm{c}}\nabla_{\mathrm{d}}\big(f_{\mathcal{Q}}\mathcal{T}^{\mathrm{cd}}\big)\right.\\\nonumber
&-&\left.\mathbb{L}_{\mathcal{M}}f_{\mathcal{T}}\right\}g_{\mathrm{a}\mathrm{b}}
-\frac{1}{2}\Box\big(f_{\mathcal{Q}}\mathcal{T}_{\mathrm{a}\mathrm{b}}\big)-2f_{\mathcal{Q}}\mathcal{R}_{\mathrm{c}(\mathrm{a}}
\mathcal{T}_{\mathrm{b})}^{\mathrm{c}}+\nabla_{\mathrm{c}}\nabla_{(\mathrm{a}}[\mathcal{T}_{\mathrm{b})}^{\mathrm{c}}
f_{\mathcal{Q}}]\\\label{g4}
&-&\big(g_{\mathrm{a}\mathrm{b}}\Box-\nabla_{\mathrm{a}}\nabla_{\mathrm{b}}\big)f_{\mathcal{R}}+2\big(f_{\mathcal{Q}}\mathcal{R}^{\mathrm{cd}}
+f_{\mathcal{T}}g^{\mathrm{cd}}\big)\frac{\partial^2\mathbb{L}_{\mathcal{M}}}{\partial
g^{\mathrm{a}\mathrm{b}}\partial g^{\mathrm{cd}}},
\end{eqnarray}
where $f_{\mathcal{R}}=\frac{\partial
f(\mathcal{R},\mathcal{T},\mathcal{Q})}{\partial
\mathcal{R}}$,~$f_{\mathcal{T}}=\frac{\partial
f(\mathcal{R},\mathcal{T},\mathcal{Q})}{\partial \mathcal{T}}$ and
$f_{\mathcal{Q}}=\frac{\partial
f(\mathcal{R},\mathcal{T},\mathcal{Q})}{\partial \mathcal{Q}}$.
Also, the mathematical expression of the D'Alambert operator is
$\Box\equiv
\frac{1}{\sqrt{-g}}\partial_\mathrm{a}\big(\sqrt{-g}g^{\mathrm{a}\mathrm{b}}\partial_{\mathrm{b}}\big)$
and $\nabla_\mathrm{c}$ indicates the covariant derivative.
Generally, the matter Lagrangian can be taken in terms of energy
density or pressure, thus we consider it as
$\mathbb{L}_{\mathcal{M}}=-\mu$ for the case of anisotropic matter
distribution which results in
$\frac{\partial^2\mathbb{L}_{\mathcal{M}}} {\partial
g^{\mathrm{a}\mathrm{b}}\partial g^{\mathrm{cd}}}=0$ \cite{22}.

To discuss the collapse of dynamical cylinder, we take line element
representing the interior geometry as
\begin{equation}\label{g6}
ds^2=-\mathcal{A}^2dt^2+\mathcal{B}^2dr^2+\mathcal{C}^2d\phi^2+dz^2,
\end{equation}
where $\mathcal{A}=\mathcal{A}(t,r)$ and
$\mathcal{B}=\mathcal{B}(t,r)$ are dimensionless, while
$\mathcal{C}=\mathcal{C}(t,r)$ has the dimension of $r$. The
$\mathrm{EMT}$ portraying anisotropic dissipative fluid is given as
\begin{align}\nonumber
\mathcal{T}_{\mathrm{a}\mathrm{b}}&=\big(\mu+\mathrm{P}_{\mathrm{r}}\big)\mathcal{U}_{\mathrm{a}}\mathcal{U}_{\mathrm{b}}
+\mathrm{P}_{\mathrm{r}}g_{\mathrm{a}\mathrm{b}}+\big(\mathrm{P}_{\phi}-\mathrm{P}_{\mathrm{r}}\big)
\mathcal{K}_{\mathrm{a}}\mathcal{K}_{\mathrm{b}}+\big(\mathrm{P}_{\mathrm{z}}-\mathrm{P}_{\mathrm{r}}\big)
\mathcal{S}_{\mathrm{a}}\mathcal{S}_{\mathrm{b}}\\\label{g5}
&+\varsigma_{\mathrm{a}}\mathcal{U}_\mathrm{b}
+\varsigma_{\mathrm{b}}\mathcal{U}_\mathrm{a}-\big(g_{\mathrm{a}\mathrm{b}}+\mathcal{U}_{\mathrm{a}}\mathcal{U}_{\mathrm{b}}\big)\alpha\Omega,
\end{align}
where $\mathrm{P}_{\mathrm{r}},~\mathrm{P}_{\phi}$ and
$\mathrm{P}_{\mathrm{z}}$ are the principal pressures and $\mu$ is
the energy density. Also, $\alpha$ and $\Omega$ are the coefficient
of bulk viscosity and the expansion scalar, respectively. The four
velocity ($\mathcal{U}_{\mathrm{a}}$), four vectors
($\mathcal{K}_{\mathrm{a}}$ and $\mathcal{S}_{\mathrm{a}}$), heat
flux $\varsigma_{\mathrm{a}}$ and $\Omega$ are defined as
\begin{align}
\mathcal{U}_{\mathrm{a}}=-\mathcal{A}\delta_{\mathrm{a}}^{0},\quad
\mathcal{K}_{\mathrm{a}}=\mathcal{C}\delta_{\mathrm{a}}^{2},\quad
\mathcal{S}_{\mathrm{a}}=\delta_{\mathrm{a}}^{3},\quad
\varsigma_{\mathrm{a}}=\varsigma\mathcal{B}\delta_{\mathrm{a}}^{1},\quad
\Omega=\mathcal{U}^{\mathrm{a}}_{;\mathrm{a}},
\end{align}
satisfying the following relations
\begin{align}
\mathcal{U}_{\mathrm{a}}\mathcal{U}^{\mathrm{a}}=-1,\quad
\mathcal{K}_{\mathrm{a}}\mathcal{K}^{\mathrm{a}}=1,\quad
\mathcal{S}_{\mathrm{a}} \mathcal{S}^{\mathrm{a}}=1,\quad
\mathcal{U}_{\mathrm{a}}
\mathcal{K}^{\mathrm{a}}=0=\mathcal{S}_{\mathrm{a}}\mathcal{K}^{\mathrm{a}}=\mathcal{U}_{\mathrm{a}}
\mathcal{S}^{\mathrm{a}}.
\end{align}
Due to the interaction of matter components and geometry in this
extended theory, the $\mathrm{EMT}$ has non-disappearing divergence,
i.e., $\nabla_\mathrm{a} \mathcal{T}^{\mathrm{a}\mathrm{b}}\neq 0$.
This exerts an extra force in the gravitational field that triggers
the non-geodesic motion of test particles. Consequently, we have
\begin{align}\nonumber
\nabla^\mathrm{a}\mathcal{T}_{\mathrm{a}\mathrm{b}}&=\frac{2}{2f_\mathcal{T}+\mathcal{R}f_\mathcal{Q}+16\pi}\bigg[\nabla_\mathrm{a}
\big(f_\mathcal{Q}\mathcal{R}^{\mathrm{c}\mathrm{a}}\mathcal{T}_{\mathrm{c}\mathrm{b}}\big)+\nabla_\mathrm{b}\big(\mathbb{L}_\mathcal{M}f_\mathcal{T}\big)
-\mathcal{G}_{\mathrm{a}\mathrm{b}}\nabla^\mathrm{a}\big(f_\mathcal{Q}\mathbb{L}_\mathcal{M}\big)\\\label{g4a}
&-\frac{1}{2}\nabla_\mathrm{b}\mathcal{T}^{\mathrm{cd}}\big(f_\mathcal{T}g_{\mathrm{cd}}+f_\mathcal{Q}\mathcal{R}_{\mathrm{cd}}\big)
-\frac{1}{2}\big\{\nabla^{\mathrm{a}}(\mathcal{R}f_{\mathcal{Q}})+2\nabla^{\mathrm{a}}f_{\mathcal{T}}\big\}\mathcal{T}_{\mathrm{a}\mathrm{b}}\bigg].
\end{align}
The trace of modified field equations is given by
\begin{align}\nonumber
&3\nabla^{\mathrm{c}}\nabla_{\mathrm{c}}
f_\mathcal{R}-\mathcal{T}(f_\mathcal{T}+8\pi)+\mathcal{R}\left(f_\mathcal{R}-\frac{\mathcal{T}}{2}f_\mathcal{Q}\right)+\frac{1}{2}
\nabla^{\mathrm{c}}\nabla_{\mathrm{c}}(f_\mathcal{Q}\mathcal{T})\\\nonumber
&+\nabla_\mathrm{a}\nabla_\mathrm{c}(f_\mathcal{Q}\mathcal{T}^{\mathrm{a}\mathrm{c}})-2f+(\mathcal{R}f_\mathcal{Q}+4f_\mathcal{T})\mathbb{L}_\mathcal{M}
+2\mathcal{R}_{\mathrm{a}\mathrm{c}}\mathcal{T}^{\mathrm{a}\mathrm{c}}f_\mathcal{Q}\\\nonumber
&-2g^{\mathrm{b}\mathrm{d}}\left(f_\mathcal{T}g^{\mathrm{a}\mathrm{c}}+f_\mathcal{Q}\mathcal{R}^{\mathrm{a}\mathrm{c}}\right)
\frac{\partial^2\mathbb{L}_\mathcal{M}}{\partial
g^{\mathrm{b}\mathrm{d}}\partial g^{\mathrm{a}\mathrm{c}}}=0.
\end{align}
The disappearance of $f_\mathcal{Q}$ from the field equations
provides the gravitational effects of $f(\mathcal{R},\mathcal{T})$
theory, whereas the $f(\mathcal{R})$ gravity can be achieved for
$f_\mathcal{T}=0=f_\mathcal{Q}$.

We adopt a standard model of the form
\begin{equation}\label{g5d}
f(\mathcal{R},\mathcal{T},\mathcal{Q})=f_1(\mathcal{R})+f_2(\mathcal{T})+f_3(\mathcal{Q})=\mathcal{R}
+\Phi\sqrt{\mathcal{T}}+\Psi\mathcal{Q}.
\end{equation}
It is noteworthy that the gravitational model produces physically
acceptable results by taking different choices of the model
parameters (involving in that model) within their noticed range. For
$\Phi=0$, this model was used to analyze isotropic systems and some
acceptable values of $\Psi$ have been acquired for which the systems
show stable behavior \cite{22a,22b}. The quantities
$\mathcal{R},~\mathcal{T}$ and $\mathcal{Q}$ of the model
\eqref{g5d} become
\begin{align}\nonumber
\mathcal{R}&=-\frac{2}{\mathcal{A}^3\mathcal{B}^3\mathcal{C}}\bigg[\mathcal{A}^3\mathcal{B}\mathcal{C}''
-\mathcal{A}\mathcal{B}^3\ddot{\mathcal{C}}-\mathcal{A}\mathcal{B}^2\mathcal{C}\ddot{\mathcal{B}}+\mathcal{A}^2\mathcal{B}\mathcal{C}\mathcal{A}''
-\mathcal{A}^3\mathcal{B}'\mathcal{C}'+\mathcal{B}^3\dot{\mathcal{A}}\dot{\mathcal{C}}\\\nonumber
&+\mathcal{A}^2\mathcal{B}\mathcal{A}'\mathcal{C}'-\mathcal{A}\mathcal{B}^2\dot{\mathcal{B}}\dot{\mathcal{C}}
+\mathcal{B}^2\mathcal{C}\dot{\mathcal{A}}\dot{\mathcal{B}}-\mathcal{A}^2\mathcal{C}\mathcal{A}'\mathcal{B}'\bigg],\\\nonumber
\mathcal{T}&=-\mu+\mathrm{P}_{\mathrm{r}}+\mathrm{P}_{\phi}+\mathrm{P}_{\mathrm{z}}-3\alpha\Omega,\\\nonumber
\mathcal{Q}&=-\frac{1}{\mathcal{A}^3\mathcal{B}^3\mathcal{C}}\bigg[\mu\big\{\mathcal{A}\mathcal{B}^2\mathcal{C}\ddot{\mathcal{B}}
-\mathcal{A}^2\mathcal{B}\mathcal{C}\mathcal{A}''+\mathcal{A}^2\mathcal{CA}'\mathcal{B}'-\mathcal{A}^2\mathcal{BA}'\mathcal{C}'
+\mathcal{AB}^3\ddot{\mathcal{C}}\\\nonumber
&-\mathcal{B}^2\mathcal{C}\dot{\mathcal{A}}\dot{\mathcal{B}}-\mathcal{B}^3\dot{\mathcal{A}}\dot{\mathcal{C}}\big\}+2\varsigma
\mathcal{AB}\big\{\mathcal{AB}\dot{\mathcal{C}}'-\mathcal{BA}'\dot{\mathcal{C}}-\mathcal{A}\dot{\mathcal{B}}\mathcal{C}'\big\}
+\big(\mathrm{P}_{\mathrm{r}}-\alpha\Omega\big)\\\nonumber
&\big\{\mathcal{B}^2\mathcal{C}\dot{\mathcal{A}}\dot{\mathcal{B}}-\mathcal{AB}^2\mathcal{C}\ddot{\mathcal{B}}+\mathcal{A}^2\mathcal{BCA}''
-\mathcal{AB}^2\dot{\mathcal{B}}\dot{\mathcal{C}}-\mathcal{A}^3\mathcal{B}'\mathcal{C}'-\mathcal{A}^2\mathcal{CA}'\mathcal{B}'\\\nonumber
&+\mathcal{A}^3\mathcal{BC}''\big\}+\big(\mathrm{P}_{\phi}-\alpha\Omega\big)\big\{\mathcal{A}^3\mathcal{BC}''
-\mathcal{A}^3\mathcal{B}'\mathcal{C}'+\mathcal{A}^2\mathcal{BA}'\mathcal{C}'-\mathcal{AB}^2\dot{\mathcal{B}}\dot{\mathcal{C}}\\\nonumber
&-\mathcal{AB}^3\ddot{\mathcal{C}}+\mathcal{B}^3\dot{\mathcal{A}}\dot{\mathcal{C}}\big\}\bigg],
\end{align}
where $.=\frac{\partial}{\partial t}$ and
$'=\frac{\partial}{\partial r}$.

The corresponding field equations are
\begin{align}\label{g8}
\frac{1}{1+\Psi\mu}\left(8\pi\mu+\frac{\mu^{(D)}}{\mathcal{A}^{2}}\right)&=
\frac{\mathcal{B}'\mathcal{C}'}{\mathcal{B}^{3}\mathcal{C}}-\frac{\mathcal{C}''}{\mathcal{B}^{2}\mathcal{C}}
+\frac{\dot{\mathcal{B}}\dot{\mathcal{C}}}{\mathcal{A}^{2}\mathcal{BC}},\\\label{g8a}
\frac{1}{1+\Psi\mu}\left(8\pi\mathrm{P}_{\mathrm{r}}-\zeta\Omega+\frac{\mathrm{P}_{\mathrm{r}}^{(D)}}{\mathcal{B}^{2}}\right)&=
\frac{\dot{\mathcal{A}}\dot{\mathcal{C}}}{\mathcal{A}^{3}\mathcal{C}}-\frac{\ddot{\mathcal{C}}}{\mathcal{A}^{2}\mathcal{C}}
+\frac{\mathcal{A}'\mathcal{C}'}{\mathcal{A}\mathcal{B}^{2}\mathcal{C}},\\\label{g8b}
\frac{1}{1+\Psi\mu}\left(8\pi\mathrm{P}_{\phi}-\zeta\Omega+\frac{\mathrm{P}_{\phi}^{(D)}}{\mathcal{C}^{2}}\right)&=
\frac{\dot{\mathcal{A}}\dot{\mathcal{B}}}{\mathcal{A}^{3}\mathcal{B}}+\frac{\mathcal{A}''}{\mathcal{A}\mathcal{B}^{2}}
-\frac{\ddot{\mathcal{B}}}{\mathcal{A}^{2}\mathcal{B}}-\frac{\mathcal{A}'\mathcal{B}'}{\mathcal{A}\mathcal{B}^{3}},\\\nonumber
\frac{1}{1+\Psi\mu}\left(8\pi\mathrm{P}_{\mathrm{z}}-\zeta\Omega+\mathrm{P}_{\mathrm{z}}^{(D)}\right)&=
\frac{\dot{\mathcal{A}}\dot{\mathcal{C}}}{\mathcal{A}^{3}\mathcal{C}}-\frac{\ddot{\mathcal{B}}}{\mathcal{A}^{2}\mathcal{B}}
-\frac{\ddot{\mathcal{C}}}{A^{2}\mathcal{C}}+\frac{\dot{\mathcal{A}}\dot{\mathcal{B}}}{\mathcal{A}^{3}\mathcal{B}}
+\frac{\mathcal{A}''}{\mathcal{A}\mathcal{B}^{2}}\\\label{g8c}
&+\frac{\mathcal{C}''}{\mathcal{B}^{2}\mathcal{C}}+\frac{\mathcal{A}'\mathcal{C}'}{\mathcal{A}\mathcal{B}^{2}\mathcal{C}}
-\frac{\mathcal{A}'\mathcal{B}'}{\mathcal{A}\mathcal{B}^{3}}-\frac{\mathcal{C}'\mathcal{B}'}{\mathcal{B}^{3}\mathcal{C}}
-\frac{\dot{\mathcal{B}}\dot{\mathcal{C}}}{\mathcal{A}^{2}\mathcal{B}\mathcal{C}},
\\\label{g8d}\frac{1}{1+\Psi\mu}\left(8\pi\varsigma-\frac{\varsigma^{(D)}}{\mathcal{AB}}\right)&=\frac{\dot{\mathcal{C}}'}{\mathcal{A}
\mathcal{B}\mathcal{C}}-\frac{\dot{\mathcal{B}}\mathcal{C}'}{\mathcal{A}\mathcal{B}^{2}\mathcal{C}}
-\frac{\mathcal{A}'\dot{\mathcal{C}}}{\mathcal{A}^{2}\mathcal{B}\mathcal{C}},
\end{align}
where $\zeta=8\pi\alpha$. These equations describe how gravity and
matter components bend spacetime. The second term on the left hand
side of the above equations
$\big(\mu^{(D)},~\mathrm{P}_{\mathrm{r}}^{(D)},~\mathrm{P}_{\phi}^{(D)},~\mathrm{P}_{\mathrm{z}}^{(D)}$
and $\varsigma^{(D)}\big)$ appear due to the modification of gravity
and their values are provided in Appendix \textbf{A}. The quantities
$\left(8\pi\mu+\frac{\mu^{(D)}}{\mathcal{A}^{2}}\right)$,
$\bigg(8\pi\mathrm{P}_{\mathrm{r}}-\zeta\Omega+\frac{\mathrm{P}_{\mathrm{r}}^{(D)}}{\mathcal{B}^{2}}\bigg)$,
$\bigg(8\pi\mathrm{P}_{\phi}-\zeta\Omega+\frac{\mathrm{P}_{\phi}^{(D)}}{\mathcal{C}^{2}}\bigg)$,
$\left(8\pi\mathrm{P}_{\mathrm{z}}-\zeta\Omega+\mathrm{P}_{\mathrm{z}}^{(D)}\right)$
and
$\left(8\pi\varsigma-\frac{\varsigma^{(D)}}{\mathcal{AB}}\right)$
depict the effective energy density, effective principal pressures
and the effective heat flux, respectively.

The C-energy within the interior geometrical structure \eqref{g6}
can be determined as \cite{41ba}
\begin{equation}\label{g13}
\tilde{m}(t,r)=\mathfrak{L}\hat{\mathrm{E}}=\frac{\mathfrak{L}}{8}(1-\mathfrak{L}^{-2}\nabla_\mathrm{a}
\hat{r}\nabla^\mathrm{a} \hat{r}),
\end{equation}
where $\hat {\mathrm{E}}$ is the total gravitational energy per
specific length of the cylinder and $\hat r=\varrho\mathfrak{L}$
symbolizes the circumference radius. The terms $\varrho$ and
$\mathfrak{L}$ are the areal radius of the cylinder and specific
length, respectively, whose mathematical expressions are
$\varrho^2=\eta_{(1)\mathrm{b}}\eta_{(1)}^{\mathrm{b}}$ and
$\mathfrak{L}^2=\eta_{(2)\mathrm{b}}\eta_{(2)}^{\mathrm{b}}$. Also,
the Killing vectors are defined as
$\eta_{(1)}=\frac{\partial}{\partial \phi}$,
$\eta_{(2)}=\frac{\partial}{\partial z}$. Equation \eqref{g13},
after some manipulation, yields the mass as
\begin{equation}\label{g14}
\tilde{m}=\frac{\mathfrak{L}}{8}\bigg[1-\bigg(\frac{\mathcal{C}'}{\mathcal{B}}\bigg)^2
+\bigg(\frac{\dot{\mathcal{C}}}{\mathcal{A}}\bigg)^2\bigg].
\end{equation}
The 3D hypersurface $\Sigma$ splits the geometry into interior and
exterior regions. The interior region is defined in Eq.\eqref{g6}
while the exterior spacetime is taken as
\begin{equation}
ds^{2}=\frac{2\mathcal{M}(v)}{R}dv^{2}-2dvdR+R^{2}\left(d\phi^{2}+\lambda^{2}dz^{2}\right),
\end{equation}
where $v$ is the retarded time. Also, $\mathcal{M}$ and $R$
symbolize the mass and radius of the exterior region. We utilize
Darmois junction conditions \cite{41bc} whose fundamental forms are
given by
\begin{itemize}
\item The continuity of the metric
coefficients of the interior and exterior spacetimes holds at the
hypersurface.
\item There is no difference between the extrinsic curvature corresponding to both geometries at $\Sigma$ that equals
the radial pressure to the heat flux for the case of dynamical fluid
distribution.
\end{itemize}

Since the collapse of a self-gravitating object is associated with
the matter sector, thus we only require to employ the second
fundamental form that yields, after some manipulation, in this
modified theory as
\begin{align}
\mathcal{M}-\tilde{m} {^\Sigma_=}\frac{\mathfrak{L}}{8}, \quad
8\pi\mathrm{P}_{\mathrm{r}}-\zeta\Omega+\frac{\mathrm{P}_{\mathrm{r}}^{(D)}}{\mathcal{B}^2}
{^\Sigma_=}8\pi\varsigma-\frac{\varsigma^{(D)}}{\mathcal{AB}}.
\end{align}
It is observed that the least satisfactory definition of the
C-energy produces the difference between masses of both regions by
$\frac{\mathfrak{L}}{8}$, which disappears in the case of spherical
spacetime. The other equation equals the effective radial pressure
and effective heat flux at the boundary $\Sigma$. This guarantees
the fulfillment of the condition of vanishing radial pressure at the
boundary only if the heat flux along with modified corrections
disappears, i.e.,
$8\pi\varsigma-\frac{\varsigma^{(D)}}{\mathcal{AB}}=0$.

\section{Dynamics of the Cylindrical Star}

Initially, Misner and Sharp formulated some dynamical quantities to
study the evolution of spherical geometry. The proper radial and
temporal derivatives were used to compute the velocity and
acceleration of the considered collapsing source. These equations
have later been used in the study of spherical as well as
cylindrical spacetimes \cite{41bb}. The dynamical equations in this
scenario are
\begin{align}\label{g21}
\mathcal{T}_{\mathrm{a};\mathrm{b}}^{(\mathrm{EFF})\mathrm{b}}\mathcal{U}^{\mathrm{a}}&=\big(8\pi\mathcal{T}_{\mathrm{a}}^{\mathrm{b}}
+\mathcal{T}_{\mathrm{a}}^{(D)\mathrm{b}}\big)_{;\mathrm{b}}\mathcal{U}^{\mathrm{a}}=0,\\\label{g22}
\mathcal{T}_{\mathrm{a};\mathrm{b}}^{(\mathrm{EFF})\mathrm{b}}\varsigma^{\mathrm{a}}&=\big(8\pi\mathcal{T}_{\mathrm{a}}^{\mathrm{b}}
+\mathcal{T}_{\mathrm{a}}^{(D)\mathrm{b}}\big)_{;\mathrm{b}}\varsigma^{\mathrm{a}}=0.
\end{align}
Equations \eqref{g21} and \eqref{g22} yield, respectively, as
\begin{align}\nonumber
&\frac{1}{\mathcal{A}^2}\bigg(8\pi\mu+\frac{\mu^{(D)}}{\mathcal{A}^2}\bigg)^.+\frac{\dot{\mathcal{B}}}{\mathcal{A}^2\mathcal{B}}
\bigg(8\pi\mu+\frac{\mu^{(D)}}{\mathcal{A}^2}+8\pi\mathrm{P}_{\mathrm{r}}
-\zeta\Omega+\frac{\mathrm{P}_{\mathrm{r}}^{(D)}}{\mathcal{\mathcal{B}}^{2}}\bigg)\\\nonumber
&+\frac{1}{\mathcal{AB}}\bigg(8\pi\varsigma-\frac{\varsigma^{(D)}}{\mathcal{AB}}\bigg)'
+\frac{\dot{\mathcal{C}}}{\mathcal{A}^2\mathcal{C}}\bigg(8\pi\mu+\frac{\mu^{(D)}}{\mathcal{A}^2}+8\pi\mathrm{P}_{\phi}-\zeta\Omega
+\frac{\mathrm{P}_{\phi}^{(D)}}{\mathcal{C}^{2}}\bigg)\\\label{g23}
&+\frac{1}{\mathcal{AB}}\bigg(8\pi\varsigma-\frac{\varsigma^{(D)}}{\mathcal{AB}}\bigg)
\bigg(\frac{2\mathcal{A}'}{\mathcal{A}}+\frac{\mathcal{C}'}{\mathcal{C}}\bigg)=0,\\\nonumber
&\frac{\mathcal{B}}{\mathcal{A}}\bigg(8\pi\varsigma-\frac{\varsigma^{(D)}}{\mathcal{AB}}\bigg)^.+\frac{\mathcal{A}'}{\mathcal{A}}\bigg(8\pi\mu
+\frac{\mu^{(D)}}{\mathcal{A}^2}+8\pi\mathrm{P}_{\mathrm{r}}-\zeta\Omega+\frac{\mathrm{P}_{\mathrm{r}}^{(D)}}{\mathcal{B}^{2}}\bigg)\\\nonumber
&+\bigg(8\pi\mathrm{P}_{\mathrm{r}}-\zeta\Omega+\frac{\mathrm{P}_{\mathrm{r}}^{(D)}}{\mathcal{B}^{2}}\bigg)'
+\frac{\mathcal{C}'}{\mathcal{C}}\bigg(8\pi\mathrm{P}_{\mathrm{r}}+\frac{\mathrm{P}_{\mathrm{r}}^{(D)}}{\mathcal{B}^{2}}-8\pi\mathrm{P}_{\phi}
-\frac{\mathrm{P}_{\phi}^{(D)}}{\mathcal{C}^{2}}\bigg)\\\label{g24}
&+\frac{\mathcal{B}}{\mathcal{A}}\bigg(8\pi\varsigma-\frac{\varsigma^{(D)}}{\mathcal{AB}}\bigg)\bigg(\frac{2\dot{\mathcal{B}}}{\mathcal{B}}
+\frac{\dot{\mathcal{C}}}{\mathcal{C}}\bigg)=0.
\end{align}
These equations play significant role in the study of variations
arising in the stellar evolution. Our goal is to discuss the
dynamics of the collapsing source, thus the definitions of proper
radial as well as temporal derivatives are \cite{29a,41bb}
\begin{equation}\label{g25}
\mathfrak{D}_{\mathrm{r}}=\frac{1}{\mathcal{C}^{\prime}}
\frac{\partial}{\partial \mathrm{r}}, \quad
\mathfrak{D}_{\mathrm{t}}=\frac{1}{\mathcal{A}}
\frac{\partial}{\partial \mathrm{t}}.
\end{equation}

The radius of an astrophysical object decreases continuously during
the collapse as gravity dominates the outward pressure.
Consequently, the velocity of interior fluid turns out to be
negative, i.e.,
\begin{equation}\label{g26}
\mathbb{U}=\mathfrak{D}_{\mathrm{t}}(\mathcal{C})=\frac{\dot{\mathcal{C}}}{\mathcal{A}}<0.
\end{equation}
Using this equation in C-energy \eqref{g14}, we have
\begin{equation}\label{g27}
\frac{\mathcal{C}'}{\mathcal{B}}=\left(1+\mathbb{U}^{2}-\frac{8
\tilde{m}}{\mathfrak{L}}\right)^{\frac{1}{2}}=\omega.
\end{equation}
The C-energy of the current cylindrical configuration yields after
applying the definition of $\mathfrak{D}_{\mathrm{t}}$ as
\begin{align}\nonumber
\mathfrak{D}_{\mathrm{t}}(\tilde{m})&=-\frac{\mathcal{C}\mathfrak{L}}{4(1+\Psi\mu)}\left\{\left(8\pi\mathrm{P}_{\mathrm{r}}-\zeta
\Omega+\frac{\mathrm{P}_{\mathrm{r}}^{(D)}}{\mathcal{B}^{2}}\right)\mathbb{U}\right.\\\label{g28}
&\left.+\left(8\pi\varsigma-\frac{\varsigma^{(D)}}{\mathcal{AB}}\right)\omega\right\},
\end{align}
which demonstrates that how the total energy varies with time. This
equation also indicates how the collapsing phenomenon is influenced
by the radial pressure, the expansion scalar as well as heat flux
and modified corrections. As $\mathbb{U}$ is negative, thus the
factor $\left(8\pi\mathrm{P}_{\mathrm{r}}-\zeta
\Omega+\frac{\mathrm{P}_{\mathrm{r}}^{(D)}}{\mathcal{B}^{2}}\right)\mathbb{U}$
on the right hand side of the above equation becomes positive, which
guarantees that the total energy of the system increases. The other
entity
$\left(8\pi\varsigma-\frac{\varsigma^{(D)}}{\mathcal{AB}}\right)\omega$
confirms the reduction of total energy as heat dissipates from the
source.

Next, in order to discuss the variation of energy between the
adjoining cylindrical surfaces, we employ the definition of
$\mathfrak{D}_{\mathrm{r}}$ on Eq.\eqref{g14} and combine it with
Eqs.\eqref{g8} and \eqref{g8d} as
\begin{align}\label{g29}
\mathfrak{D}_{\mathrm{r}}(\tilde{m})&=\frac{\mathcal{C}\mathfrak{L}}{4(1+\Psi\mu)}\left\{\left(8\pi\mu+\frac{\mu^{(D)}}{\mathcal{A}^{2}}\right)
+\left(8\pi\varsigma-\frac{\varsigma^{(D)}}{\mathcal{AB}}\right)\frac{\mathbb{U}}{\omega}\right\}.
\end{align}
The collapse rate of the current setup is also affected by the
effective energy density. The term
$\left(8\pi\mu+\frac{\mu^{(D)}}{\mathcal{A}^{2}}\right)$ in the
above equation ultimately increases the total energy of the matter
source. The next entity,
$\left(8\pi\varsigma-\frac{\varsigma^{(D)}}{\mathcal{AB}}\right)\frac{\mathbb{U}}{\omega}$,
reveals that the heat energy dissipates from the system, as the
fluid has negative velocity. The acceleration of the collapsing
source can be calculated by taking proper temporal derivative of
$\mathbb{U}$ as
\begin{eqnarray}\label{g30}
\mathfrak{D}_{\mathrm{t}}(\mathbb{U})=-\frac{\mathcal{C}}{1+\Psi\mu}\bigg(8\pi\mathrm{P}_{\mathrm{r}}-\zeta
\Omega+\frac{\mathrm{P}_{\mathrm{r}}^{(D)}}{\mathcal{B}^{2}}\bigg)-\frac{\tilde{m}}{\mathcal{C}^2}+\frac{\omega\mathcal{A}'}{\mathcal{AB}}
+\frac{\mathfrak{L}}{8\mathcal{C}^{2}}\big(1+\mathbb{U}^{2}-\omega^{2}\big).
\end{eqnarray}
One can get the value of $\frac{\mathcal{A}'}{\mathcal{A}}$ from
Eq.\eqref{g24} as
\begin{align}\nonumber
\frac{\mathcal{A}'}{\mathcal{A}}&=-\frac{1}{\left(8\pi\mu+\frac{\mu^{(D)}}{\mathcal{A}^2}+8\pi\mathrm{P}_{\mathrm{r}}-\zeta\Omega
+\frac{\mathrm{P}_{\mathrm{r}}^{(D)}}{\mathcal{B}^{2}}\right)}\bigg\{\frac{\mathcal{B}}{\mathcal{A}}\bigg(8\pi\varsigma
-\frac{\varsigma^{(D)}}{\mathcal{AB}}\bigg)^.\\\nonumber
&+\bigg(8\pi\mathrm{P}_{\mathrm{r}}-\zeta\Omega+\frac{\mathrm{P}_{\mathrm{r}}^{(D)}}{\mathcal{B}^{2}}\bigg)'
+\frac{\mathcal{C}'}{\mathcal{C}}\bigg(8\pi\mathrm{P}_{\mathrm{r}}+\frac{\mathrm{P}_{\mathrm{r}}^{(D)}}{\mathcal{B}^{2}}-8\pi\mathrm{P}_{\phi}
-\frac{\mathrm{P}_{\phi}^{(D)}}{\mathcal{C}^{2}}\bigg)\\\label{g31}
&+\frac{\mathcal{B}}{\mathcal{A}}\bigg(8\pi\varsigma-\frac{\varsigma^{(D)}}{\mathcal{AB}}\bigg)\bigg(\frac{2\dot{\mathcal{B}}}{\mathcal{B}}
+\frac{\dot{\mathcal{C}}}{\mathcal{C}}\bigg)\bigg\}.
\end{align}
Inserting this value in Eq.\eqref{g30}, we have
\begin{align}\nonumber
&\mathfrak{D}_{\mathrm{t}}(\mathbb{U})\bigg(8\pi\mu+\frac{\mu^{(D)}}{\mathcal{A}^2}+8\pi\mathrm{P}_{\mathrm{r}}-\zeta\Omega
+\frac{\mathrm{P}_{\mathrm{r}}^{(D)}}{\mathcal{B}^{2}}\bigg)=-\bigg\{\frac{\tilde{m}}{\mathcal{C}^{2}}
-\frac{\mathfrak{L}}{8\mathcal{C}^{2}}\bigg(1+\mathbb{U}^{2}\bigg)\\\nonumber
&+\frac{\mathcal{C}}{1+\Psi\mu}\bigg(8\pi\mathrm{P}_{\mathrm{r}}-\zeta\Omega
+\frac{\mathrm{P}_{\mathrm{r}}^{(D)}}{\mathcal{B}^{2}}\bigg)\bigg\}\bigg(8\pi\mu+\frac{\mu^{(D)}}{\mathcal{A}^2}+8\pi\mathrm{P}_{\mathrm{r}}
-\zeta\Omega+\frac{\mathrm{P}_{\mathrm{r}}^{(D)}}{\mathcal{B}^{2}}\bigg)\\\nonumber
&-\frac{\omega^{2}}{\mathcal{C}}\bigg\{\bigg(8\pi\mathrm{P}_{\mathrm{r}}+\frac{\mathrm{P}_{\mathrm{r}}^{(D)}}{\mathcal{B}^{2}}
-8\pi\mathrm{P}_{\phi}-\frac{\mathrm{P}_{\phi}^{(D)}}{\mathcal{C}^{2}}\bigg)+\frac{\mathfrak{L}}{8
\mathcal{C}}\bigg(8\pi\mu+\frac{\mu^{(D)}}{\mathcal{A}^2}+8\pi\mathrm{P}_{\mathrm{r}}-\zeta\Omega\\\nonumber
&+\frac{\mathrm{P}_{\mathrm{r}}^{(D)}}{\mathcal{B}^{2}}\bigg)\bigg\}-\omega\bigg\{\frac{1}{B}\bigg(8\pi\mathrm{P}_{\mathrm{r}}
-\zeta\Omega+\frac{\mathrm{P}_{\mathrm{r}}^{(D)}}{\mathcal{B}^{2}}\bigg)'
+\mathfrak{D}_{\mathrm{t}}\bigg(8\pi\varsigma-\frac{\varsigma^{(D)}}{\mathcal{AB}}\bigg)+\frac{1}{\mathcal{A}}\\\label{g32}
&\times\bigg(8\pi\varsigma-\frac{\varsigma^{(D)}}{\mathcal{AB}}\bigg)\bigg(\frac{2\dot{\mathcal{B}}}{\mathcal{B}}
+\frac{\dot{\mathcal{C}}}{\mathcal{C}}\bigg)\bigg\}.
\end{align}

The left side of this equation represents the Newtonian force as the
product of acceleration ($\mathfrak{D}_{\mathrm{t}}\mathbb{U}$) and
the term
$\bigg(8\pi\mu+\frac{\mu^{(D)}}{\mathcal{A}^2}+8\pi\mathrm{P}_{\mathrm{r}}-\zeta\Omega
+\frac{\mathrm{P}_{\mathrm{r}}^{(D)}}{\mathcal{B}^{2}}\bigg)$
(refers to the inertial mass density) appears. On the other hand,
the same term also arises in the first term on the right side, that
now presents the gravitational mass density. Thus, the equivalence
of these both masses leads to the fulfillment of the equivalence
principle. The second curly bracket determines the impact of
gravitational mass density and effective stresses in $\mathrm{r}$ as
well as $\phi$ directions on the collapse rate. The role of gradient
of effective radial pressure and the expansion scalar in this
scenario can be manifested through the entity
$\bigg(8\pi\mathrm{P}_{\mathrm{r}}
-\zeta\Omega+\frac{\mathrm{P}_{\mathrm{r}}^{(D)}}{\mathcal{B}^{2}}\bigg)'$.
Also, the last two terms containing the heat flux and modified
corrections can well describe hydrodynamics of the cylinder.

\section{Transport Equations}

As the $\mathrm{EMT}$ \eqref{g5} involves the heat flux, the
transport equations in this regard are very useful tool to analyze
the structural evolution of compact geometry. They also disclose how
some physical quantities such as mass, heat and momentum are
evaluated during the collapse. The diffusion process is supported by
the following transport equation given as
\begin{equation}\label{g33}
\varrho\mathrm{h}^{\mathrm{a}\mathrm{b}}\mathcal{U}^{\mathrm{c}}\bar{\varsigma}_{\mathrm{b};\mathrm{c}}+\bar{\varsigma}^{\mathrm{a}}=
-\eta\mathrm{h}^{\mathrm{a}\mathrm{b}}\left(\tau_{,\mathrm{b}}+\tau\mathrm{a}_{\mathrm{b}}\right)-\frac{1}{2}\eta\tau^{2}\left(\frac{\varrho
\mathcal{U}^{\mathrm{b}}}{\eta\tau^{2}}\right)_{;\mathrm{b}}\bar{\varsigma}^{\mathrm{a}},
\end{equation}
where
$\bar{\varsigma}=\left(8\pi\varsigma-\frac{\varsigma^{(D)}}{\mathcal{AB}}\right)$
and
$\mathrm{h}^{\mathrm{a}\mathrm{b}}=g^{\mathrm{a}\mathrm{b}}+\mathcal{U}^{\mathrm{a}}
\mathcal{U}^{\mathrm{b}}$ is the projection tensor. Also,
$\eta,~\varrho,~\tau$ and $\mathrm{a}_{\mathrm{b}}$ \big(i.e.,
$\mathrm{a}_1=\frac{\mathcal{A}'}{\mathcal{A}}$\big) are
mathematical symbols of the thermal conductivity, relaxation time,
temperature and acceleration, respectively. After some
simplification, Eq.\eqref{g33} produces
\begin{align}\nonumber
\mathcal{B}
\mathfrak{D}_{\mathrm{t}}\left(8\pi\varsigma-\frac{\varsigma^{(D)}}{\mathcal{AB}}\right)&=-\frac{\eta\tau'}{\varrho}
-\frac{\eta\tau}{\varrho}\left(\frac{\mathcal{A}'}{\mathcal{A}}\right)-\frac{\eta\tau^{2}\mathcal{B}}{2\mathcal{A}\varrho}
\left(\frac{\varrho}{\eta\tau^{2}}\right)^.\left(8\pi\varsigma-\frac{\varsigma^{(D)}}{\mathcal{AB}}\right)\\\label{g34}
&-\frac{\mathcal{B}}{2\mathcal{A}}\left(\frac{3\dot{\mathcal{B}}}{\mathcal{B}}+\frac{\dot{\mathcal{C}}}{\mathcal{C}}
+\frac{2\mathcal{A}}{\varrho}\right)\left(8\pi\varsigma-\frac{\varsigma^{(D)}}{\mathcal{AB}}\right).
\end{align}
This equation becomes after inserting the value of
$\frac{\mathcal{A}'}{\mathcal{A}}$ as
\begin{align}\nonumber
\mathcal{B}
\mathfrak{D}_{\mathrm{t}}\left(8\pi\varsigma-\frac{\varsigma^{(D)}}{\mathcal{AB}}\right)&=
-\frac{\eta\tau^{2}\mathcal{B}}{2\mathcal{A}\varrho}
\left(\frac{\varrho}{\eta\tau^{2}}\right)^.\left(8\pi\varsigma-\frac{\varsigma^{(D)}}{\mathcal{AB}}\right)
-\frac{\mathcal{B}}{2\mathcal{A}}\left(8\pi\varsigma-\frac{\varsigma^{(D)}}{\mathcal{AB}}\right)\\\nonumber
&\times\bigg(\frac{3\dot{\mathcal{B}}}{\mathcal{B}}+\frac{\dot{\mathcal{C}}}{\mathcal{C}}
+\frac{2\mathcal{A}}{\varrho}\bigg)-\frac{\eta\tau\mathcal{B}}{\varrho\omega}\bigg\{\mathfrak{D}_{\mathrm{t}}(\mathbb{U})
-\frac{\mathfrak{L}}{8\mathcal{C}^{2}}\big(1+\mathbb{U}^{2}-\omega^{2}\big)\\\label{g35}
&+\frac{\tilde{m}}{\mathcal{C}^2}+\frac{\mathcal{C}}{1+\Psi\mu}\bigg(8\pi\mathrm{P}_{\mathrm{r}}-\zeta\Omega
+\frac{\mathrm{P}_{\mathrm{r}}^{(D)}}{\mathcal{B}^{2}}\bigg)\bigg\}-\frac{\eta\tau'}{\varrho},
\end{align}
which demonstrates how much variation takes place in heat energy
with the passage of time. This equation also explains the impact of
temperature, thermal conductivity and relaxation time on the
self-gravitating systems. Eliminating
$\mathfrak{D}_{\mathrm{t}}\left(8\pi\varsigma-\frac{\varsigma^{(D)}}{\mathcal{AB}}\right)$
from Eqs.\eqref{g32} and \eqref{g35}, we obtain
\begin{align}\nonumber
&\mathfrak{D}_{\mathrm{t}}(\mathbb{U})\bigg(8\pi\mu+\frac{\mu^{(D)}}{\mathcal{A}^2}+8\pi\mathrm{P}_{\mathrm{r}}-\zeta\Omega
+\frac{\mathrm{P}_{\mathrm{r}}^{(D)}}{\mathcal{B}^{2}}-\frac{\eta\tau}{\varrho}\bigg)=
-\bigg\{\frac{\tilde{m}}{\mathcal{C}^{2}}-\frac{\mathfrak{L}}{8\mathcal{C}^{2}}\bigg(1+\mathbb{U}^{2}\bigg)\\\nonumber
&+\frac{\mathcal{C}}{1+\Psi\mu}\bigg(8\pi\mathrm{P}_{\mathrm{r}}-\zeta\Omega
+\frac{\mathrm{P}_{\mathrm{r}}^{(D)}}{\mathcal{B}^{2}}\bigg)\bigg\}\bigg(8\pi\mu+\frac{\mu^{(D)}}{\mathcal{A}^2}+8\pi\mathrm{P}_{\mathrm{r}}
-\zeta\Omega+\frac{\mathrm{P}_{\mathrm{r}}^{(D)}}{\mathcal{B}^{2}}-\frac{\eta\tau}{\varrho}\bigg)\\\nonumber
&\times\bigg\{1-\frac{\eta\tau}{\varrho}\bigg(8\pi\mu+\frac{\mu^{(D)}}{\mathcal{A}^2}+8\pi\mathrm{P}_{\mathrm{r}}-\zeta\Omega
+\frac{\mathrm{P}_{\mathrm{r}}^{(D)}}{\mathcal{B}^{2}}\bigg)^{-1}\bigg\}
+\omega^{2}\bigg\{\frac{\eta\tau\mathfrak{L}}{8\varrho\mathcal{C}^{2}}-\frac{\mathfrak{L}}{8\mathcal{C}^{2}}\\\nonumber
&\times\bigg(8\pi\mu+\frac{\mu^{(D)}}{\mathcal{A}^{2}}+8\pi\mathrm{P}_{\mathrm{r}}-\zeta\Omega
+\frac{\mathrm{P}_{\mathrm{r}}^{(D)}}{\mathcal{B}^{2}}\bigg)-\frac{1}{\mathcal{C}}\bigg(8\pi\mathrm{P}_{\mathrm{r}}
+\frac{\mathrm{P}_{\mathrm{r}}^{(D)}}{\mathcal{B}^{2}}-8\pi\mathrm{P}_{\phi}-\frac{\mathrm{P}_{\phi}^{(D)}}{\mathcal{C}^{2}}\bigg)\bigg\}
\\\nonumber & -\omega\bigg[-\frac{\eta\tau'}{\varrho\mathcal{B}}+\frac{1}{\mathcal{B}}\bigg(8\pi\mathrm{P}_{\mathrm{r}}-\zeta\Omega
+\frac{\mathrm{P}_{\mathrm{r}}^{(D)}}{\mathcal{B}^{2}}\bigg)^{\prime}+\frac{1}{2}\bigg\{\frac{\dot{\mathcal{B}}}{\mathcal{AB}}
+\frac{\dot{\mathcal{C}}}{\mathcal{AC}}-\frac{2}{\varrho}
-\frac{\eta\tau^2}{\mathcal{A}\varrho}\bigg(\frac{\varrho}{\eta\tau^2}\bigg)^.\bigg\}\\\label{g36}
&\times\bigg(8\pi\varsigma-\frac{\varsigma^{(D)}}{\mathcal{AB}}\bigg)\bigg].
\end{align}
We can rearrange this equation as
\begin{align}\nonumber
&\mathfrak{D}_{\mathrm{t}}(\mathbb{U})\bigg(8\pi\mu+\frac{\mu^{(D)}}{\mathcal{A}^2}+8\pi\mathrm{P}_{\mathrm{r}}-\zeta\Omega
+\frac{\mathrm{P}_{\mathrm{r}}^{(D)}}{\mathcal{B}^{2}}\bigg)\big(1-\mathfrak{H}\big)=
-\mathcal{F}_{\mathrm{grav}}\big(1-\mathfrak{H}\big)\\\nonumber
&+\mathcal{F}_{\mathrm{hyd}}+\omega^{2}\bigg\{\frac{\eta\tau\mathfrak{L}}{8\varrho\mathcal{C}^{2}}-\frac{\mathfrak{L}}{8\mathcal{C}^{2}}
\bigg(8\pi\mu+\frac{\mu^{(D)}}{\mathcal{A}^{2}}+8\pi\mathrm{P}_{\mathrm{r}}-\zeta\Omega
+\frac{\mathrm{P}_{\mathrm{r}}^{(D)}}{\mathcal{B}^{2}}\bigg)\\\label{g37}
&-\frac{1}{\mathcal{C}}\bigg(8\pi\mathrm{P}_{\mathrm{r}}+\frac{\mathrm{P}_{\mathrm{r}}^{(D)}}{\mathcal{B}^{2}}
-8\pi\mathrm{P}_{\phi}-\frac{\mathrm{P}_{\phi}^{(D)}}{\mathcal{C}^{2}}\bigg)\bigg\},
\end{align}
where
\begin{align}\label{g38}
\mathfrak{H}=&\frac{\eta\tau}{\varrho}\bigg(8\pi\mu+\frac{\mu^{(D)}}{\mathcal{A}^2}+8\pi\mathrm{P}_{\mathrm{r}}-\zeta\Omega
+\frac{\mathrm{P}_{\mathrm{r}}^{(D)}}{\mathcal{B}^{2}}\bigg)^{-1},\\\nonumber
\mathcal{F}_{\mathrm{grav}}=&\bigg(8\pi\mu+\frac{\mu^{(D)}}{\mathcal{A}^2}+8\pi\mathrm{P}_{\mathrm{r}}-\zeta\Omega
+\frac{\mathrm{P}_{\mathrm{r}}^{(D)}}{\mathcal{B}^{2}}\bigg)
\bigg\{\frac{\tilde{m}}{\mathcal{C}^{2}}-\frac{\mathfrak{L}}{8\mathcal{C}^{2}}\left(\mathbb{U}^{2}+1\right)\\\label{g39}
&+\frac{\mathcal{C}}{1+\Psi\mu}\bigg(8\pi\mathrm{P}_{\mathrm{r}}-\zeta\Omega
+\frac{\mathrm{P}_{\mathrm{r}}^{(D)}}{\mathcal{B}^{2}}\bigg)\bigg\},\\\nonumber
\mathcal{F}_{\mathrm{hyd}}=&-\omega\bigg[\frac{1}{\mathcal{B}}\bigg(8\pi\mathrm{P}_{\mathrm{r}}-\zeta\Omega
+\frac{\mathrm{P}_{\mathrm{r}}^{(D)}}{\mathcal{B}^{2}}\bigg)^{\prime}
+\frac{1}{2}\bigg\{\frac{\dot{\mathcal{B}}}{\mathcal{AB}}+\frac{\dot{\mathcal{C}}}{\mathcal{AC}}-\frac{2}{\varrho}
-\frac{\eta\tau^2}{\mathcal{A}\varrho}\bigg(\frac{\varrho}{\eta\tau^2}\bigg)^.\bigg\}\\\label{g40}
&\times\bigg(8\pi\varsigma-\frac{\varsigma^{(D)}}{\mathcal{AB}}\bigg)-\frac{\eta\tau'}{\varrho\mathcal{B}}\bigg].
\end{align}
Equation \eqref{g37} explains how the collapse rate is affected by
different forces, comprising Newtonian
($\mathcal{F}_{\mathrm{newtn}}$), hydrodynamical
($\mathcal{F}_{\mathrm{hyd}}$) and gravitational
($\mathcal{F}_{\mathrm{grav}}$) forces. It is known that energy
always dissipates (in the form of radiation, convection and
conduction) from higher to lower energy state of the system. The
energy of a star is dissipated through radiations, if photons
acquire it from the higher phase of that object. On the other hand,
when photons do not possess all the energy, it will be dissipated by
convection. The hot gasses in this phenomenon move to the upper zone
and thus radiate energy, whereas cooler gases attain energy by
traveling towards the hot zone. There occur continuous collisions of
atoms inside an object due to which every atom transfers its energy
to the nearest one, and thus energy dissipates in the form of
conduction.

Equation \eqref{g37} involves an entity $(1-\mathfrak{H})$ that
acknowledges the equivalence principle, while the gravitational mass
density and the term $(\mathfrak{H})$ \big(defined in
Eq.\eqref{g38}\big) are inversely proportional to each other. This
relation provides the fact that the gravitational force and the
quantity $(1-\mathfrak{H})$ are strongly affected by each other,
leading to the following different cases.
\begin{itemize}
\item The entity $(1-\mathfrak{H})$ remains positive only for $\mathfrak{H}<1$,
which results in the negative gravitational force (i.e., repulsive
force) due to the appearance of minus sign in the first term on the
right side of Eq.\eqref{g37}. Consequently, the collapse rate
diminishes.

\item The rate of the cylindrical collapse increases for the case when $\mathfrak{H}>1$,
i.e., $(1-\mathfrak{H})<0$.

\item Finally, if we consider $\mathfrak{H}=1$, the gravitational as well as inertial
forces disappear and we have from Eq.\eqref{g37} as
\begin{align}\nonumber
&\omega^{2}\bigg\{\frac{\eta\tau\mathfrak{L}}{8\varrho\mathcal{C}^{2}}-\frac{\mathfrak{L}}{8\mathcal{C}^{2}}
\bigg(8\pi\mu+\frac{\mu^{(D)}}{\mathcal{A}^{2}}+8\pi\mathrm{P}_{\mathrm{r}}-\zeta\Omega
+\frac{\mathrm{P}_{\mathrm{r}}^{(D)}}{\mathcal{B}^{2}}\bigg)\\\nonumber
&-\frac{1}{\mathcal{C}}\bigg(8\pi\mathrm{P}_{\mathrm{r}}+\frac{\mathrm{P}_{\mathrm{r}}^{(D)}}{\mathcal{B}^{2}}
-8\pi\mathrm{P}_{\phi}-\frac{\mathrm{P}_{\phi}^{(D)}}{\mathcal{C}^{2}}\bigg)\bigg\}=\omega\bigg[-\frac{\eta\tau'}{\varrho\mathcal{B}}\\\nonumber
&+\frac{1}{\mathcal{B}}\bigg(8\pi\mathrm{P}_{\mathrm{r}}-\zeta\Omega+\frac{\mathrm{P}_{\mathrm{r}}^{(D)}}{\mathcal{B}^{2}}\bigg)^{\prime}
+\frac{1}{2}\bigg(8\pi\varsigma-\frac{\varsigma^{(D)}}{\mathcal{AB}}\bigg)\\\label{g41}
&\times\bigg\{\frac{\dot{\mathcal{B}}}{\mathcal{AB}}+\frac{\dot{\mathcal{C}}}{\mathcal{AC}}-\frac{2}{\varrho}
-\frac{\eta\tau^2}{\mathcal{A}\varrho}\bigg(\frac{\varrho}{\eta\tau^2}\bigg)^.\bigg\}\bigg].
\end{align}
\end{itemize}
This equation expresses the involvement of temperature, thermal
conductivity, the bulk viscosity and the modified corrections in the
collapsing phenomenon. The equilibrium position of the collapsing
cylinder is supported by the hydrodynamical force (given on left
side of the above equation), and hence, the collapse rate is
reduced.

\section{Relation between the Weyl Scalar and Effective Physical Quantities}

In this section, we develop some relations between effective
physical variables and the Weyl scalar
($\mathbb{C}^2=\mathfrak{C}_{\mathrm{c}\mathrm{a}\mathrm{d}\mathrm{b}}\mathfrak{C}^{\mathrm{c}\mathrm{a}\mathrm{d}\mathrm{b}}$,
where $\mathfrak{C}_{\mathrm{c}\mathrm{a}\mathrm{d}\mathrm{b}}$ is
the Weyl tensor). This scalar can be expressed as a linear
combination of the Kretchmann scalar
($\mathbb{R}=\mathcal{R}_{\mathrm{c}\mathrm{a}\mathrm{d}\mathrm{b}}\mathcal{R}^{\mathrm{c}\mathrm{a}\mathrm{d}\mathrm{b}},~\mathcal{R}_{\mathrm{c}\mathrm{a}\mathrm{d}\mathrm{b}}$
is the Riemann tensor), the Ricci tensor
($\mathcal{R}_{\mathrm{a}\mathrm{b}}$) and the Ricci scalar as
\cite{35c}
\begin{align}\label{g42}
\mathbb{C}^2=\mathbb{R}-2\mathcal{R}_{\mathrm{a}\mathrm{b}}\mathcal{R}^{\mathrm{a}\mathrm{b}}+\frac{1}{3}\mathcal{R}^2.
\end{align}
The scalar $\mathbb{R}$ can be manipulated as
\begin{align}\label{g43}
\mathbb{R}=\frac{4}{\mathcal{A}^4\mathcal{B}^4\mathcal{C}^4}\left\{\mathcal{C}^4\big(\mathcal{R}^{0101}\big)^2
+\mathcal{B}^4\big(\mathcal{R}^{0202}\big)^2+\mathcal{A}^4\big(\mathcal{R}^{1212}\big)^2
-\frac{\mathcal{A}^2\mathcal{B}^2\big(\mathcal{R}^{1202}\big)^2}{2}\right\}.
\end{align}
For the considered spacetime \eqref{g6}, the Ricci scalar, non-zero
components of the Riemann tensor and the Ricci tensor in terms of
the Einstein tensor are
\begin{align}\nonumber
\mathcal{R}&=-2\bigg(\frac{\mathcal{G}_{11}}{\mathcal{B}^2}+\frac{\mathcal{G}_{22}}{\mathcal{C}^2}
-\frac{\mathcal{G}_{00}}{\mathcal{A}^2}\bigg),\\\nonumber
\mathcal{R}^{0101}&=\frac{\mathcal{G}_{22}}{\big(\mathcal{ABC}\big)^2},\quad
\mathcal{R}^{0202}=\frac{\mathcal{G}_{11}}{\big(\mathcal{ABC}\big)^2},\quad
\mathcal{R}^{1212}=\frac{\mathcal{G}_{00}}{\big(\mathcal{ABC}\big)^2},\\\nonumber
\mathcal{R}^{0212}&=\frac{\mathcal{G}_{01}}{\big(\mathcal{ABC}\big)^2},\quad
\mathcal{R}_{00}=\mathcal{A}^2\bigg(\frac{\mathcal{G}_{11}}{\mathcal{B}^2}+\frac{\mathcal{G}_{22}}{\mathcal{C}^2}\bigg),\quad
\mathcal{R}_{01}=\mathcal{G}_{01},\\\nonumber
\mathcal{R}_{11}&=\mathcal{B}^2\bigg(\frac{\mathcal{G}_{00}}{\mathcal{A}^2}-\frac{\mathcal{G}_{22}}{\mathcal{C}^2}\bigg),\quad
\mathcal{R}_{22}=\mathcal{C}^2\bigg(\frac{\mathcal{G}_{00}}{\mathcal{A}^2}-\frac{\mathcal{G}_{11}}{\mathcal{B}^2}\bigg).
\end{align}
These values provide the scalar $\mathbb{R}$ \eqref{g43} as
\begin{align}\label{g44}
\mathbb{R}=\frac{4}{\mathcal{A}^4\mathcal{B}^4\mathcal{C}^4}\left\{\mathcal{B}^4\mathcal{C}^4\mathcal{G}_{00}^2
+\mathcal{A}^4\mathcal{C}^4\mathcal{G}_{11}^2+\mathcal{A}^4\mathcal{B}^4\mathcal{G}_{22}^2
-4\mathcal{A}^2\mathcal{B}^2\mathcal{C}^4\mathcal{G}_{01}^2\right\}.
\end{align}
Inserting these equations in Eq.\eqref{g42}, the Weyl scalar takes
the form as
\begin{align}\nonumber
\mathbb{C}^2&=\frac{4}{3\mathcal{A}^4\mathcal{B}^4\mathcal{C}^4}\big\{\mathcal{B}^4\mathcal{C}^4\mathcal{G}_{00}^2
+\mathcal{A}^4\mathcal{C}^4\mathcal{G}_{11}^2+\mathcal{A}^4\mathcal{B}^4\mathcal{G}_{22}^2
+\mathcal{A}^2\mathcal{B}^2\mathcal{C}^4\mathcal{G}_{00}\mathcal{G}_{11}\\\label{g45}
&+\mathcal{A}^2\mathcal{B}^4\mathcal{C}^2\mathcal{G}_{00}\mathcal{G}_{22}
-\mathcal{A}^4\mathcal{B}^2\mathcal{C}^2\mathcal{G}_{11}\mathcal{G}_{22}\big\}.
\end{align}
Using this equation and the field equations \eqref{g8}-\eqref{g8b}
yields
\begin{align}\nonumber
\frac{\sqrt{3}\mathbb{C}}{2}&=\bigg[\bigg\{\frac{1}{1+\Psi\beta}\bigg(8\pi\mu+\frac{\mu^{(D)}}{\mathcal{A}^{2}}+8\pi\mathrm{P}_{\mathrm{r}}
+\frac{\mathrm{P}_{\mathrm{r}}^{(D)}}{\mathcal{B}^{2}}-8\pi\mathrm{P}_{\phi}
-\frac{\mathrm{P}_{\phi}^{(D)}}{\mathcal{C}^{2}}\bigg)\bigg\}^2\\\nonumber
&-\frac{1}{1+\Psi\beta}\bigg\{\bigg(8\pi\mathrm{P}_{\mathrm{r}}-\zeta\Omega+\frac{\mathrm{P}_{\mathrm{r}}^{(D)}}{\mathcal{B}^{2}}\bigg)
\bigg(8\pi\mathrm{P}_{\phi}-\zeta\Omega+\frac{\mathrm{P}_{\phi}^{(D)}}{\mathcal{C}^{2}}\bigg)\\\label{g46}
&+\bigg(8\pi\mathrm{P}_{\mathrm{r}}+\frac{\mathrm{P}_{\mathrm{r}}^{(D)}}{\mathcal{B}^{2}}+2\zeta\Omega-24\pi\mathrm{P}_{\phi}
-\frac{3\mathrm{P}_{\phi}^{(D)}}{\mathcal{C}^{2}}\bigg)\bigg(8\pi\mu+\frac{\mu^{(D)}}{\mathcal{A}^{2}}\bigg)\bigg\}\bigg]^{\frac{1}{2}}.
\end{align}

The necessary and sufficient condition for a spacetime to be
conformally flat is the energy density homogeneity. We check the
validity of this result in the present modified gravity. For this,
we have considered the standard model \eqref{g5d}. We take the case
when $\mathcal{R}=\mathcal{R}_0$ and $f_2(\mathcal{T})$ as well as
$f_3(\mathcal{Q})$ are treated as constants, thus Eq.\eqref{g46} is
left with
\begin{align}\nonumber
\frac{\sqrt{3}\mathbb{C}}{2}&=\bigg[\bigg\{8\pi\bigg(\mu+\mathrm{P}_{\mathrm{r}}-\mathrm{P}_{\phi}-\frac{\mathcal{C}_{0}}{16\pi}\bigg)\bigg\}^2
-\bigg(8\pi\mathrm{P}_{\mathrm{r}}+2\zeta\Omega-24\pi\mathrm{P}_{\phi}-\mathcal{C}_{0}\bigg)\\\label{g47}
&\times\bigg(8\pi\mu-\frac{\mathcal{C}_{0}}{2}\bigg)-\bigg(8\pi\mathrm{P}_{\mathrm{r}}-\zeta\Omega+\frac{\mathcal{C}_{0}}{2}\bigg)
\bigg(8\pi\mathrm{P}_{\phi}-\zeta\Omega+\frac{\mathcal{C}_{0}}{2}\bigg)\bigg]^{\frac{1}{2}},
\end{align}
where
$\mathcal{C}_{0}=\Phi\sqrt{\mathcal{T}_{0}}+\Psi\mathcal{Q}_{0}$ is
a constant term. This equation depicts that inhomogeneity in the
energy density of the fluid is induced due to the presence of the
bulk viscosity and the principal pressures. The above relation also
says that inhomogeneity in the system (during evolution) is
increased by the tidal forces \cite{36a}. The only possibility to
obtain conformally flat spacetime is the consideration of dust
matter distribution, that gives in the absence of bulk viscosity as
\begin{align}\label{g48}
\sqrt{3}\mathbb{C}&=16\pi\bigg(\mu-\frac{\mathcal{C}_{0}}{16\pi}\bigg)
\quad \Rightarrow \quad \sqrt{3}\mathbb{C}'=16\pi\mu'.
\end{align}
We observe from this equation that homogeneous energy density (i.e.,
$\mu'=0$) implies conformally flat spacetime ($\mathbb{C}=0$ through
regular axis condition) and vice-versa, and hence the required
condition is obtained.

\section{Conclusions}

Our cosmos comprises an abundance of astronomical systems whose
structural formation is highly influenced by an appealing
phenomenon, named as the gravitational collapse. The study of
gravitational waves through multiple observations has prompted
several astrophysicists to investigate the collapsing rate of
self-gravitating geometries in $\mathbb{GR}$ and other extended
theories \cite{44}. This article is based on the formulation of
dynamical description of the cylindrical fluid distribution to
investigate the changes that are gradually produced within the
system in the background of
$f(\mathcal{R},\mathcal{T},\mathcal{R}_{\mathrm{ab}}\mathcal{T}^{\mathrm{ab}})$
gravity. The effect of heat dissipation and principal pressures
(such as $\mathrm{P}_{\mathrm{r}},~\mathrm{P}_{\phi}$ and
$\mathrm{P}_{\mathrm{z}}$) on the interior geometry has been
considered. We have formulated two dynamical equations through
Misner-Sharp formalism to examine the variations in the total energy
with respect to radial as well as temporal coordinates.

We have constructed the transport equation as well as some
fundamental forces (such as the gravitational, hydrodynamical and
Newtonian) and then coupled them with dynamical equations to analyze
the impact of modified gravity on the collapse rate. The entity
$\mathfrak{H}$ is found to be in direct relation with temperature as
well as thermal conductivity and inversely related with
gravitational mass density. In the following, we summarize our
results.
\begin{itemize}
\item The entity $\mathfrak{H}$ will be less as compared to $\mathbb{GR}$ for positive
effect of the modified corrections. This leads to the increment in
the term $(1-\mathfrak{H})$ as well as the gravitational force.
However, the appearance of minus sign ultimately diminishes the
collapse rate.

\item The rate of cylindrical collapse may increase for the case
when the effect of correction terms is negative.

\item One cannot say anything about the decrement/increment in the
collapsing rate when the corrections of this modified gravity
involved in $\mathfrak{H}$ have opposite signs.
\end{itemize}
The relevance of density inhomogeneity and the Weyl scalar has also
been developed. By implying some constraints on the considered
modified model, it has been shown that the homogenous density and
the conformal flatness of the current setup imply each other. It is
worth mentioning here that the tidal forces involving in the Weyl
tensor produce more inhomogeneity in the fluid during the
evolutionary process. All these results can be recovered in
$\mathbb{GR}$ for $\Phi=0=\Psi$.

\section*{Appendix A}

\renewcommand{\theequation}{A\arabic{equation}}
\setcounter{equation}{0} The modified corrections in
Eqs.\eqref{g8}-\eqref{g8d} are
\begin{align}\nonumber
\mu^{(D)}&=-\frac{\mathcal{A}^2\big(\Phi\sqrt{\mathcal{T}}+\Psi\mathcal{Q}\big)}{2}+\Psi\bigg\{\mu\bigg(\frac{4\dot{\mathcal{A}}^2}{\mathcal{A}^2}
-\frac{\mathcal{A}'^2}{\mathcal{B}^2}+\frac{\mathcal{AA}''}{\mathcal{B}^2}+\frac{3\dot{\mathcal{A}}\dot{\mathcal{B}}}{\mathcal{AB}}
-\frac{\mathcal{AA}'\mathcal{B}'}{\mathcal{B}^3}\\\nonumber
&+\frac{3\dot{\mathcal{A}}\dot{\mathcal{C}}}{\mathcal{AC}}+\frac{\mathcal{AA}'\mathcal{C}'}{\mathcal{B}^2\mathcal{C}}
-\frac{2\ddot{\mathcal{C}}}{\mathcal{C}}-\frac{2\ddot{\mathcal{B}}}{\mathcal{B}}\bigg)+\dot{\mu}\bigg(\frac{\dot{\mathcal{C}}}{2\mathcal{C}}
+\frac{\dot{\mathcal{B}}}{2\mathcal{B}}\bigg)-\mu'\bigg(\frac{2\mathcal{AA}'}{\mathcal{B}^2}
-\frac{\mathcal{A}^2\mathcal{B}'}{2\mathcal{B}^3}\\\nonumber
&+\frac{\mathcal{A}^2\mathcal{C}'}{2\mathcal{B}^2\mathcal{C}}\bigg)-\frac{\mu''\mathcal{A}^2}{2\mathcal{B}^2}
+\mathrm{P}_\mathrm{r}\bigg(\frac{4\mathcal{A}^2\mathcal{B}'^2}{\mathcal{B}^4}-\frac{\mathcal{A}^2\mathcal{B}''}{\mathcal{B}^3}
+\frac{\dot{\mathcal{B}}^2}{\mathcal{B}^2}\bigg)-\frac{\dot{\mathrm{P}}_\mathrm{r}\dot{\mathcal{B}}}{2\mathcal{B}}
-\frac{5\mathrm{P}'_\mathrm{r}\mathcal{A}^2\mathcal{B}'}{2\mathcal{B}^3}\\\nonumber
&+\frac{\mathrm{P}''_\mathrm{r}\mathcal{A}^2}{2\mathcal{B}^2}-\mathrm{P}_{\phi}\bigg(\frac{\dot{\mathcal{C}}^2}{\mathcal{C}^2}
-\frac{\mathcal{A}^2\mathcal{C}'^2}{\mathcal{B}^2\mathcal{C}^2}\bigg)-\frac{\dot{\mathrm{P}}_{\phi}\dot{\mathcal{C}}}{2\mathcal{C}}
+\frac{\mathrm{P}'_{\phi}\mathcal{A}^2\mathcal{C}'}{2\mathcal{B}^2\mathcal{C}}-\varsigma\bigg(\frac{2\mathcal{A}\dot{\mathcal{C}}'}{\mathcal{BC}}
-\frac{2\dot{\mathcal{A}}\mathcal{B}'}{\mathcal{B}^2}\\\nonumber
&+\frac{2\dot{\mathcal{A}}'}{\mathcal{B}}-\frac{4\dot{\mathcal{A}}\mathcal{A}'}{\mathcal{AB}}-\frac{2\mathcal{A}'\dot{\mathcal{B}}}{\mathcal{B}^2}
-\frac{2\mathcal{A}\dot{\mathcal{B}}\mathcal{C}'}{\mathcal{B}^2\mathcal{C}}-\frac{2\mathcal{A}'\dot{\mathcal{C}}}{\mathcal{BC}}
-\frac{2\mathcal{A}'\dot{\mathcal{B}}}{\mathcal{B}^2}\bigg)-\frac{2\dot{\varsigma}\mathcal{A}'}{\mathcal{B}}
-\frac{2\varsigma'\dot{\mathcal{A}}}{\mathcal{B}}\bigg\},\\\nonumber
\mathrm{P}_{\mathrm{r}}^{(D)}&=\frac{\mathcal{B}^2}{2}\bigg\{\bigg(\frac{\Phi}{\sqrt{\mathcal{T}}}+\Psi\mathcal{R}\bigg)\mathrm{P}_{\mathrm{r}}
+\Phi\sqrt{\mathcal{T}}+\Psi\mathcal{Q}+\frac{\Phi\mu}{\sqrt{\mathcal{T}}}\bigg\}
+\Psi\bigg\{\mu\bigg(\frac{\ddot{\mathcal{A}}\mathcal{B}^2}{\mathcal{A}^3}-\frac{\mathcal{A}'^2}{\mathcal{A}^2}\\\nonumber
&-\frac{4\dot{\mathcal{A}}^2B^2}{\mathcal{A}^4}\bigg)+\frac{5\dot{\mu}\dot{\mathcal{A}}B^2}{2\mathcal{A}^3}
+\frac{\mu'\mathcal{A}'}{2\mathcal{A}}-\frac{\ddot{\mu}\mathcal{B}^2}{2\mathcal{A}^2}
+\mathrm{P}_\mathrm{r}\bigg(\frac{\mathcal{B}\dot{\mathcal{A}}\dot{\mathcal{B}}}{\mathcal{A}^3}-\frac{3\mathcal{A}'\mathcal{B}'}{\mathcal{AB}}
-\frac{\mathcal{B}\dot{\mathcal{B}}\dot{\mathcal{C}}}{\mathcal{A}^2C}\\\nonumber
&+\frac{\dot{\mathcal{B}}^2}{\mathcal{A}^2}-\frac{4\mathcal{B}'^2}{\mathcal{B}^2}-\frac{\mathcal{B}\ddot{\mathcal{B}}}{\mathcal{A}^2}
-\frac{3\mathcal{B}'\mathcal{C}'}{\mathcal{BC}}+\frac{2\mathcal{A}''}{\mathcal{A}}+\frac{2\mathcal{C}''}{\mathcal{C}}\bigg)
+\dot{\mathrm{P}}_\mathrm{r}\bigg(\frac{\mathcal{B}^2\dot{\mathcal{C}}}{2\mathcal{A}^2\mathcal{C}}
-\frac{\mathcal{B}^2\dot{\mathcal{A}}}{2\mathcal{A}^3}\\\nonumber
&+\frac{2\mathcal{B}\dot{\mathcal{B}}}{\mathcal{A}^2}\bigg)-\mathrm{P}'_\mathrm{r}\bigg(\frac{\mathcal{A}'}{2\mathcal{A}}
+\frac{\mathcal{C}'}{2\mathcal{C}}\bigg)+\frac{\ddot{\mathrm{P}}_\mathrm{r}\mathcal{B}^2}{2\mathcal{A}^2}
-\mathrm{P}_{\phi}\bigg(\frac{B^2\dot{\mathcal{C}}^2}{A^2\mathcal{C}^2}-\frac{\mathcal{C}'^2}{\mathcal{C}^2}\bigg)
-\frac{\dot{\mathrm{P}}_{\phi}B^2\dot{\mathcal{C}}}{2A^2\mathcal{C}}\\\nonumber
&-\frac{\mathrm{P}'_{\phi}\mathcal{C}'}{2\mathcal{C}}+\varsigma\bigg(\frac{2\dot{\mathcal{B}}'}{\mathcal{A}}
-\frac{2\dot{\mathcal{A}}\mathcal{B}'}{\mathcal{A}^2}-\frac{4\mathcal{A}'\dot{\mathcal{B}}}{\mathcal{A}^2}
-\frac{4\dot{\mathcal{B}}\mathcal{B}'}{\mathcal{AB}}+\frac{2\mathcal{B}\dot{\mathcal{C}}'}{\mathcal{AC}}
-\frac{2\dot{\mathcal{B}}\mathcal{C}'}{\mathcal{AC}}-\frac{2\mathcal{B}\mathcal{A}'\dot{\mathcal{C}}}{\mathcal{A}^2\mathcal{C}}\bigg)\\\nonumber
&+\frac{2\dot{\varsigma}\mathcal{B}'}{\mathcal{A}}+\frac{2\varsigma'\dot{\mathcal{B}}}{\mathcal{A}}\bigg\},\\\nonumber
\mathrm{P}_{\phi}^{(D)}&=\frac{\mathcal{C}^2}{2}\bigg\{\bigg(\frac{\Phi}{\sqrt{\mathcal{T}}}+\Psi\mathcal{R}\bigg)\mathrm{P}_{\phi}
+\Phi\sqrt{\mathcal{T}}+\Psi\mathcal{Q}+\frac{\Phi\mu}{\sqrt{\mathcal{T}}}\bigg\}+\Psi\bigg\{\mu\bigg(\frac{\ddot{\mathcal{A}}\mathcal{C}^2}
{\mathcal{A}^3}-\frac{\mathcal{A}'^2\mathcal{C}^2}{\mathcal{A}^2\mathcal{B}^2}\\\nonumber
&-\frac{4\dot{\mathcal{A}}^2\mathcal{C}^2}{\mathcal{A}^4}\bigg)+\frac{5\dot{\mu}\dot{\mathcal{A}}\mathcal{C}^2}{2\mathcal{A}^3}
+\frac{\mu'\mathcal{A}'\mathcal{C}^2}{2\mathcal{A}\mathcal{B}^2}-\frac{\ddot{\mu}\mathcal{C}^2}{2\mathcal{A}^2}
+\mathrm{P}_\mathrm{r}\bigg(\frac{\mathcal{B}''\mathcal{C}^2}{\mathcal{B}^3}-\frac{4\mathcal{B}'^2\mathcal{C}^2}{\mathcal{B}^4}
-\frac{\dot{\mathcal{B}}^2\mathcal{C}^2}{\mathcal{A}^2\mathcal{B}^2}\bigg)\\\nonumber
&+\frac{\dot{\mathrm{P}}_\mathrm{r}\mathcal{C}^2\dot{\mathcal{B}}}{2\mathcal{A}^2\mathcal{B}}
+\frac{5\mathrm{P}'_\mathrm{r}\mathcal{C}^2\mathcal{B}'}{2\mathcal{B}^3}-\frac{{\mathrm{P}}''_\mathrm{r}\mathcal{C}^2}{2\mathcal{B}^2}
+\mathrm{P}_{\phi}\bigg(\frac{\dot{\mathcal{C}}^2}{\mathcal{A}^2}-\frac{\mathcal{C}\ddot{\mathcal{C}}}{\mathcal{A}^2}
+\frac{\mathcal{C}\dot{\mathcal{A}}\dot{\mathcal{C}}}{\mathcal{A}^3}-\frac{\mathcal{C}\mathcal{B}'\mathcal{C}'}{\mathcal{B}^3}
-\frac{\mathcal{C}\dot{\mathcal{B}}\dot{\mathcal{C}}}{\mathcal{A}^2\mathcal{B}}\\\nonumber
&-\frac{\mathcal{C}'^2}{\mathcal{B}^2}-\frac{\mathcal{C}\mathcal{A}'\mathcal{C}'}{\mathcal{AB}^2}+\frac{\mathcal{C}\mathcal{C}''}
{\mathcal{B}^2}\bigg)+\dot{\mathrm{P}}_{\phi}\bigg(\frac{\mathcal{C}^2\dot{\mathcal{B}}}{2\mathcal{A}^2\mathcal{B}}
-\frac{\mathcal{C}^2\dot{\mathcal{A}}}{2\mathcal{A}^3}+\frac{2\mathcal{C}\dot{\mathcal{C}}}{\mathcal{A}^2}\bigg)
-\mathrm{P}'_{\phi}\bigg(\frac{\mathcal{C}^2\mathcal{A}'}{2\mathcal{AB}^2}\\\nonumber
&-\frac{\mathcal{C}^2\mathcal{B}'}{2\mathcal{B}^3}+\frac{2\mathcal{CC}'}{\mathcal{B}^2}\bigg)
+\frac{\ddot{\mathrm{P}}_{\phi}\mathcal{C}^2}{2\mathcal{A}^2}-\frac{\mathrm{P}''_{\phi}\mathcal{C}^2}{2\mathcal{B}^2}-\varsigma\bigg(
\frac{3\mathcal{C}^2\dot{\mathcal{A}}\mathcal{A}'}{\mathcal{A}^3\mathcal{B}}+\frac{3\mathcal{C}^2\dot{\mathcal{B}}\mathcal{B}'}{\mathcal{AB}^3}
+\frac{3\mathcal{C}^2\mathcal{A}'\dot{\mathcal{B}}}{\mathcal{A}^2\mathcal{B}^2}\\\nonumber
&-\frac{\mathcal{C}^2\dot{\mathcal{A}}'}{\mathcal{A}^2\mathcal{B}}-\frac{\mathcal{C}^2\dot{\mathcal{B}}'}{\mathcal{AB}^2}
+\frac{\mathcal{C}^2\dot{\mathcal{A}}\mathcal{B}'}{\mathcal{A}^2\mathcal{B}^2}\bigg)
+\dot{\varsigma}\bigg(\frac{2\mathcal{C}^2\mathcal{A}'}{\mathcal{A}^2\mathcal{B}}+\frac{\mathcal{C}^2\mathcal{B}'}{\mathcal{AB}^2}\bigg)
+\varsigma'\bigg(\frac{\mathcal{C}^2\dot{\mathcal{A}}}{\mathcal{A}^2\mathcal{B}}
+\frac{2\mathcal{C}^2\dot{\mathcal{B}}}{\mathcal{AB}^2}\bigg)\\\nonumber
&-\frac{\dot{\varsigma}'\mathcal{C}^2}{\mathcal{AB}}\bigg\},\\\nonumber
\mathrm{P}_{\mathrm{z}}^{(D)}&=\frac{1}{2}\bigg\{\bigg(\frac{\Phi}{\sqrt{\mathcal{T}}}+\Psi\mathcal{R}\bigg)\mathrm{P}_{\phi}
+\Phi\sqrt{\mathcal{T}}+\Psi\mathcal{Q}+\frac{\Phi\mu}{\sqrt{\mathcal{T}}}\bigg\}+\Psi\bigg\{\mu\bigg(\frac{\ddot{\mathcal{A}}}
{\mathcal{A}^3}-\frac{\mathcal{A}'^2}{\mathcal{A}^2\mathcal{B}^2}\\\nonumber
&-\frac{4\dot{\mathcal{A}}^2}{\mathcal{A}^4}\bigg)+\frac{5\dot{\mu}\dot{\mathcal{A}}}{2\mathcal{A}^3}
+\frac{\mu'\mathcal{A}'}{2\mathcal{A}\mathcal{B}^2}-\frac{\ddot{\mu}}{2\mathcal{A}^2}
+\mathrm{P}_\mathrm{r}\bigg(\frac{\mathcal{B}''}{\mathcal{B}^3}-\frac{4\mathcal{B}'^2}{\mathcal{B}^4}
-\frac{\dot{\mathcal{B}}^2}{\mathcal{A}^2\mathcal{B}^2}\bigg)
+\frac{\dot{\mathrm{P}}_\mathrm{r}\dot{\mathcal{B}}}{2\mathcal{A}^2\mathcal{B}}\\\nonumber
&+\frac{5\mathrm{P}'_\mathrm{r}\mathcal{B}'}{2\mathcal{B}^3}-\frac{{\mathrm{P}}''_\mathrm{r}}{2\mathcal{B}^2}
+\mathrm{P}_{\phi}\bigg(\frac{\mathcal{C}'^2}{\mathcal{B}^2\mathcal{C}^2}-\frac{\dot{\mathcal{C}}^2}{\mathcal{A}^2\mathcal{C}^2}\bigg)
+\frac{\dot{\mathrm{P}}_{\phi}\dot{\mathcal{C}}}{2\mathcal{A}^2\mathcal{C}}-\frac{\mathrm{P}'_{\phi}\mathcal{C}'}{2\mathcal{B}^2\mathcal{C}}
+\dot{\mathrm{P}}_{\mathrm{z}}\bigg(\frac{\dot{\mathcal{B}}}{2\mathcal{A}^2\mathcal{B}}\\\nonumber
&-\frac{\dot{A}}{2\mathcal{A}^3}+\frac{\dot{\mathcal{C}}}{2\mathcal{A}^2\mathcal{C}}\bigg)
+\mathrm{P}'_{\mathrm{z}}\bigg(\frac{\mathcal{B}'}{2\mathcal{B}^3}-\frac{A'}{2\mathcal{AB}^2}-\frac{\mathcal{C}'}{2\mathcal{B}^2\mathcal{C}}\bigg)
+\frac{\ddot{\mathrm{P}}_{\mathrm{z}}}{2\mathcal{A}^2}-\frac{\mathrm{P}''_{\mathrm{z}}}{2\mathcal{B}^2}\\\nonumber
&+\varsigma\bigg(\frac{\dot{\mathcal{A}}'}{\mathcal{A}^2\mathcal{B}}-\frac{3\dot{\mathcal{A}}\mathcal{A}'}{\mathcal{A}^3\mathcal{B}}
-\frac{\dot{\mathcal{A}}\mathcal{B}'}{\mathcal{A}^2\mathcal{B}^2}+\frac{\dot{\mathcal{B}}'}{\mathcal{AB}^2}
-\frac{3\mathcal{A}'\dot{\mathcal{B}}}{\mathcal{A}^2\mathcal{B}^2}-\frac{3\dot{\mathcal{B}}\mathcal{B}'}{\mathcal{AB}^3}\bigg)
+\dot{\varsigma}\bigg(\frac{2\mathcal{A}'}{\mathcal{A}^2\mathcal{B}}\\\nonumber
&+\frac{\mathcal{B}'}{\mathcal{AB}^2}\bigg)+\varsigma'\bigg(\frac{\dot{\mathcal{A}}}{\mathcal{A}^2\mathcal{B}}
+\frac{2\dot{\mathcal{B}}}{\mathcal{AB}^2}\bigg)-\frac{\dot{\varsigma}'\mathcal{C}^2}{\mathcal{AB}}\bigg\},\\\nonumber
\varsigma^{(D)}&=-\frac{\varsigma\mathcal{AB}}{2}\bigg(\frac{\Phi}{\sqrt{\mathcal{T}}}+\Psi\mathcal{R}\bigg)
+\Psi\bigg\{\mu\bigg(\frac{A'\dot{\mathcal{C}}}{\mathcal{AC}}-\frac{\dot{\mathcal{C}}'}{\mathcal{C}}
+\frac{\dot{\mathcal{B}}C'}{\mathcal{BC}}\bigg)+\frac{\dot{\mu}\mathcal{A}'}{2\mathcal{A}}+\frac{\mu'\dot{\mathcal{A}}}{2\mathcal{B}}\\\nonumber
&-\frac{\dot{\mu}'}{2}+\mathrm{P}_\mathrm{r}\bigg(\frac{\dot{\mathcal{C}}'}{\mathcal{C}}-\frac{A'\dot{\mathcal{C}}}{\mathcal{AC}}
-\frac{\dot{\mathcal{B}}C'}{\mathcal{BC}}\bigg)-\frac{\dot{\mathrm{P}}_\mathrm{r}\mathcal{A}'}{2\mathcal{A}}
-\frac{\mathrm{P}'_\mathrm{r}\dot{\mathcal{B}}}{2\mathcal{B}}+\frac{\dot{\mathrm{P}}'_\mathrm{r}}{2}
+\varsigma\bigg(\frac{2\ddot{\mathcal{B}}}{\mathcal{A}}-\frac{\ddot{\mathcal{A}}\mathcal{B}}{\mathcal{A}^2}\\\nonumber
&-\frac{4\dot{\mathcal{A}}\dot{\mathcal{B}}}{\mathcal{A}^2}+\frac{2\dot{\mathcal{A}}^2\mathcal{B}}{\mathcal{A}^3}
+\frac{\mathcal{A}'^2}{\mathcal{AB}}+\frac{4\mathcal{A}'\mathcal{B}'}{\mathcal{B}^2}
-\frac{2\mathcal{A}''}{\mathcal{B}}+\frac{\mathcal{AB}''}{\mathcal{B}^2}-\frac{\dot{\mathcal{B}}^2}{\mathcal{AB}}
+\frac{\mathcal{B}\ddot{\mathcal{C}}}{\mathcal{AC}}-\frac{\mathcal{AC}''}{\mathcal{BC}}\\\nonumber
&-\frac{2\mathcal{AB}'^2}{\mathcal{B}^3}-\frac{3\mathcal{B}\dot{\mathcal{A}}\dot{\mathcal{C}}}{2\mathcal{A}^2\mathcal{C}}
+\frac{\dot{\mathcal{B}}\dot{\mathcal{C}}}{2\mathcal{AC}}-\frac{\mathcal{A}'\mathcal{C}'}{2\mathcal{BC}}
+\frac{3\mathcal{AB}'\mathcal{C}'}{2\mathcal{B}^2\mathcal{C}}\bigg)
-\dot{\varsigma}\bigg(\frac{2\dot{\mathcal{A}}\mathcal{B}}{\mathcal{A}^2}+\frac{\mathcal{B}\dot{\mathcal{C}}}{2\mathcal{AC}}\bigg)\\\nonumber
&+\varsigma'\bigg(\frac{2\mathcal{A}\mathcal{B}'}{\mathcal{B}^2}+\frac{\mathcal{A}\mathcal{C}'}{2\mathcal{BC}}\bigg)\bigg\}.
\end{align}

\subsection*{Data Availability Statement}

This manuscript has no associated data.

\end{document}